\documentstyle[graphicx,twoside,fleqn,espcrc2]{article}

\newcommand{\beq}{\begin{equation}}
\newcommand{\eeq}{\end{equation}}        
\newcommand{\bqa}{\begin{eqnarray}}        
\newcommand{\eqa}{\end{eqnarray}}

\newcommand{\fig}[1]{{\frenchspacing Fig.~(\ref{#1})}}
\newcommand{\tab}[1]{{\frenchspacing Tab.~(\ref{#1})}}
\newcommand{\mpr}{\frac{m_{\pi}}{m_{\rho}}}
\newcommand{\sesam}{{\sf SESAM}}
\newcommand{\txl}{{\sf T$\chi$L}}

\title{
\hfill\begin{minipage}{0pt}\scriptsize\vspace*{-1.5cm} \begin{tabbing}
\hspace*{\fill} IFUM 631/FT \\
\hspace*{\fill} IFUP-TH 43/98\\
\hspace*{\fill} HLRZ1998-51\\ 
\hspace*{\fill} WUP-TH 98-28 
\end{tabbing} 
\end{minipage}\\[-8pt]
Decorrelating Topology with HMC}
     
\author{\frenchspacing 
Th. Lippert\address{Department of Physics, University of Wuppertal,
    D-42097 Wuppertal, Germany}\thanks{Talk presented by Th.~Lippert},
B. All\'es\address{Dipartimento di Fisica,
    Universit\`a di Milano and INFN, Via Celoria 16, I-20133 Milano,
    Italy},
  G. Bali\address{Institut f\"ur Physik, Humboldt Universit\"at,
    Invalidenstrasse 110, D-10115 Berlin, Germany},
  M. D'Elia\address{Dipartimento di Fisica dell'Universit\`a and INFN,
    Piazza Torricelli 2, I-56126-Pisa, Italy}$^,$\address{Department of
    Natural Sciences, University of Cyprus P.O. Box 537, Nicosia
    CY-1678, Cyprus},
A. Di Giacomo$^{\rm{d}}$,
N. Eicker\address{HLRZ, c/o FZ-J\"{u}lich, and DESY-Hamburg, D-52425
  J\"{u}lich, Germany},
S. G\"usken$^{\rm{a}}$,
K. Schilling$^{\rm{a,f}}$,
A. Spitz$^{\rm{f}}$,
T. Struckmann$^{\rm{a}}$,
P. Ueberholz$^{\rm{a}}$,
and J. Viehoff$^{\rm{a}}$\nonfrenchspacing}

\begin{document}

\begin{abstract}
  The investigation of the decorrelation efficiency of the HMC
  algorithm with respect to vacuum topology is a prerequisite for
  trustworthy full QCD simulations, in particular for the computation
  of topology sensitive quantities. We demonstrate that for
  $\mpr$-ratios $\ge 0.69$ sufficient tunneling between the
  topological sectors can be achieved, for two flavours of dynamical
  Wilson fermions close to the scaling region ($\beta=5.6$).  Our
  results are based on time series of length 5000 trajectories.
\end{abstract}

\maketitle

\section{Introduction}

Topology is a fundamental feature of continuum quantum field theories
with important bearings on elementary particle physics issues, as
there is the proton spin content, the large $\eta '$ mass or the role
of instantons for the vacuum structure.

Stochastic lattice sampling of QCD configurations must be sufficiently
ergodic in the topological sectors of phase space in order to allow
for the proper computation of topological quantities. In particular
for full QCD, the tunneling rates induced by the hybrid Monte Carlo
algorithm between the topological sectors are expected to decrease
dramatically when approaching the chiral limit, as known from {\em
  staggered fermion} simulations \cite{KURAMASHI,MMP,PISA}.  For full
QCD with dynamical {\em Wilson fermions} the actual size of the $\mpr$
window accessible by todays simulations is still an open question
which we are going to address here \cite{TOPOLOGY}.

In the last years, various techniques for the reliable computation of
topological continuum quantities from the discrete lattice data have
been put forth \cite{ALLESCRITICAL}.  In the present investigations,
we will just make use of the field theoretical definition
\cite{FABRICIUS} of the topological charge density\footnote{
  Alternatively, we follow Smit and Vink \cite{SMIT} avoiding cooling.
  With $G$ being the quark propagator we compute $Q$ via stochastic
  estimates \cite{VIEHOFF}:
\begin{equation} \label{smitvinck}
  Q = m\, \kappa_P \langle Tr(\gamma_5 G)\rangle_U.
\label{fermionic}
\end{equation}}
\begin{equation}
  Q(x) =
  \frac{g^2}{64\pi^2}\epsilon^{\mu\nu\rho\sigma}F^{a}_{\mu\nu}(x)F^{a}_{\rho\sigma}(x)
\label{gluonic}
\end{equation}
with application of cooling \cite{PISA} as it is not necessary to
compute the renormalization of $Q$.

For our investigations, we exploit the \sesam\ and \txl\ 
\cite{TSUKUBA} samples consisting of three time histories on lattices
of size $16^3\times 32$ at $\kappa=0.156$, $0.157$, and $0.1575$, at
$\mpr$ ratios of $0.839(4)$, $0.755(7)$, and $0.69(1)$, respectively
and of configurations on $24^3\times 40$ lattices, again at
$\beta=5.6$, with $\kappa=0.1575$ and $0.158$, the latter giving rise
to $\mpr\approx 0.58(2)$. The length of the samples is 5000
trajectories. For $\kappa=0.158$, the length is 3500 trajectories.
From these samples, we take sets of $N=200$ `decorrelated'
configurations.

\section{Results}

\begin{figure}[h]
  \centerline{\includegraphics[width=\columnwidth]{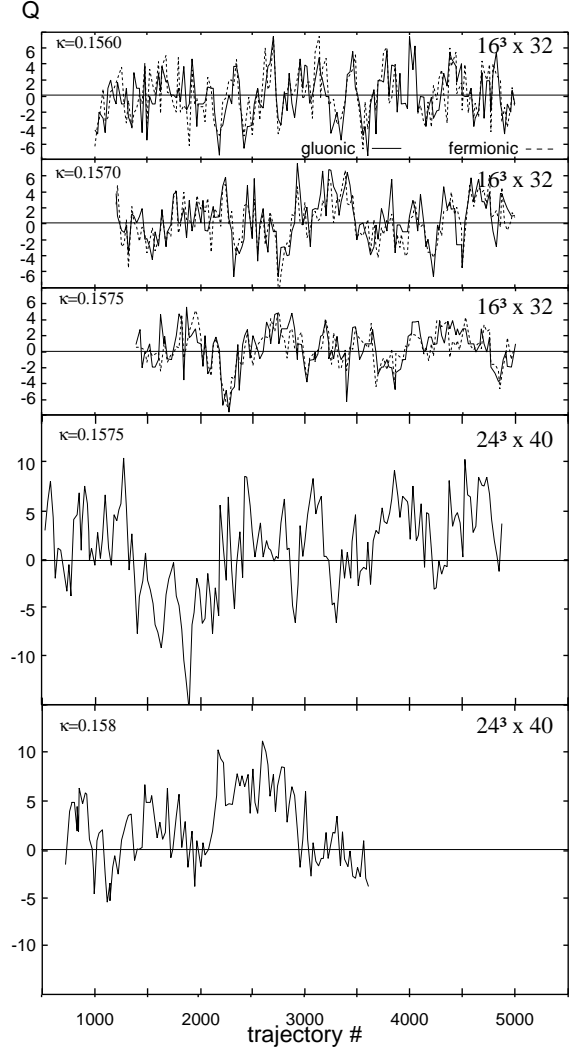}}
\caption{Time series of unrenormalized topological charge. We plot series
  from 3 \sesam\ and 2 \txl\ ensembles.
\label{FIG:5SERIES}}
\end{figure}

In \fig{FIG:5SERIES} we present the time series of the topological
charge $Q$ from the five samples. The first three canvases are
computed from \sesam\ ensembles measured on lattices of size
$16^3\times 32$, the next two are from \txl\ ensembles from
$24^3\times 40$ systems.  We analyzed every 25th configuration.  In
the first three graphs, the results from the gluonic and fermionic
measurements are super-imposed demonstrating nice agreement.  For all
three \sesam\ masses, the figure tells us that HMC can create
sufficient tunneling in the \sesam\ $\mpr$ window.

We quantify the tunneling between the topological sectors
defining the mobility for tunneling as
\begin{equation}
D_d = \frac{1}{N}\sum_{i=1}^{N}|(\tilde Q((i+1)\cdot d)- \tilde
Q(i\cdot d))|.
\label{MOBILITY}
\end{equation}
\tab{TAB:FREQUENCY} demonstrates that the mobility is decreasing when
going more chiral.  \tabcolsep3.9pt
\begin{table}[tb]
\caption{First part: mobility  $D_{25}$. Second part:
  mobility normalized by dividing through square root of the volume.
\label{TAB:FREQUENCY}}
\begin{tabular}{l||c|c|c|c|c}
  $\kappa$ & .156 & .157 & .1575 & 
  0.1580 & {quenched} \\
\hline
\multicolumn{6}{c}{$D_{25}$}\\
  \hline
  $16^3\times 32$ & 2.8  & 2.5 & 1.9 &  -  & -  \\
  \hline
  $24^3\times 40$ & -    & -   & 3.8 & 2.8  & - \\
  \hline
  $16^4$          & -    & -   &  - & - & 1.2    \\
\hline
\multicolumn{6}{c}{$D_{25}/\sqrt{V}$}\\
\hline
  $16^3\times 32$ & 2.8  & 2.5 & 1.9 &  -  & -  \\
  \hline
  $24^3\times 40$ & -    & -   & 1.9 & 1.4 & - \\
  \hline
  $16^4$          & -    & -   &  - & - & 1.7    \\
\end{tabular}
\vspace*{-.3cm}
\end{table}

With the $24^3\times 40$ system, at $\kappa=0.1575$, we can assess the
volume dependence. Note that $Q$ is an extensive quantity.  Still we
find the tunneling rates sufficient, however, with smaller quark mass,
at $\kappa=0.158$, the mobility decreases. As we expect the mobility
to scale approximately with the square root of the volume, we
normalize the numbers accordingly, see the second part of
\tab{TAB:FREQUENCY}.

\begin{figure}[htb]
\centerline{\includegraphics[width=\columnwidth]{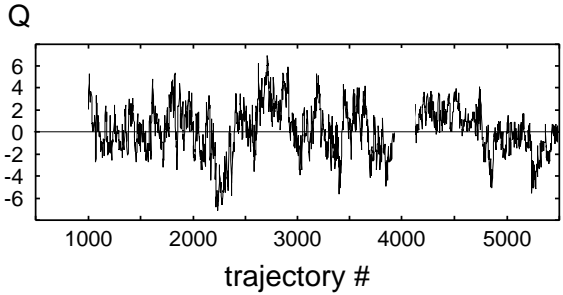}}
\caption{Fine scan of unrenormalized
  $Q$ for $\kappa=0.1575$ on the $16^3\times 32$ lattice. 
\label{FIG:FINE}}
\end{figure}

Analyzing the time history on every second trajectory (at
$\kappa=0.1575$) we find very frequent tunneling, see \fig{FIG:FINE}.
On this series, we can compute the autocorrelation function.  The
exponential autocorrelation time is $\tau_{exp}=80(10)$.  The
integrated autocorrelation time turns out to be $\tau_{int}=54(4)$,
compatible with the decorrelation of other observables as found in
\cite{SESAMAUTOLATTICE}.

\begin{figure}[!htb]
\includegraphics[width=\columnwidth]{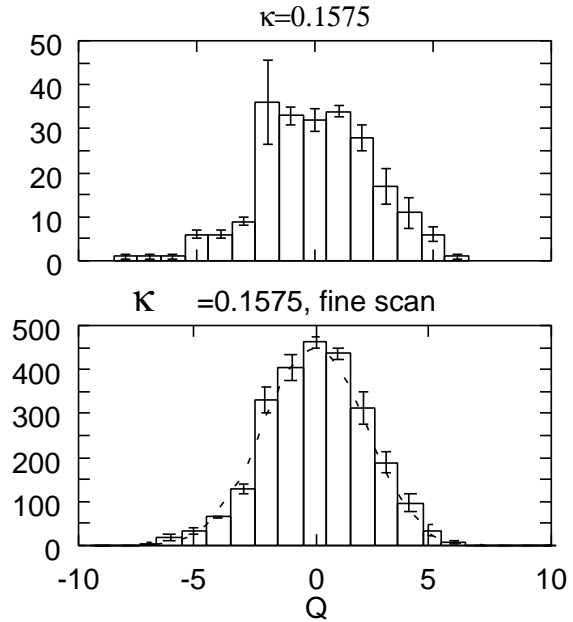}
\caption{Histograms of the unrenormalized topological charge computed from 
  Eq.~(1)
\label{FIG:4HISTOS}}
\end{figure}

Histogramming \fig{FIG:FINE} we find a smooth Gaussian, see
\fig{FIG:4HISTOS}. This is of course less pronounced for the small
decorrelated samples of size 200, see \fig{FIG:4HISTOS}, however, the
distributions are still well peaked at $Q=0$.

\section{Summary}

As the main result, we find that HMC tunnels well between the vacuum
sectors for $\mpr\ge 0.69$.  Autocorrelation times are slightly larger
than for the yet worst case observable, the minimal eigenvalue of the
Wilson fermion matrix.  Therefore, we are confident to be able to
carry out reliable computations of hadronic properties related to
topology, like the proton spin content and the topological
susceptibility for $\mpr$-values down to $0.69$.

\end{document}